# Approximations for Standard Normal Distribution Function and Its Invertible

**Omar M. Eidous  and  Mohammad Y. Al-Rawwash**

Department of Statistics

Faculty of Science

Yarmouk University

**Abstract**

In this paper, we introduce a new approximation of the cumulative distribution function of the standard normal distribution based on Tocher's approximation. Also, we assess the quality of the new approximation using two criteria namely the maximum absolute error and the mean absolute error. The approximation is expressed in closed form and it produces a maximum absolute error of $4.43 \times 10^{-10}$ while the mean absolute error is $9.62 \times 10^{-11}$. In addition, we propose an approximation of the inverse cumulative function of the standard normal distribution based on Polya approximation and compare the accuracy of our findings with some of the existing approximations. The results show that our approximations surpass other the existing ones based on the aforementioned accuracy measures.

**Key words:** Normal distribution, Approximations, Cumulative distribution function, Maximum absolute error, Mean absolute error, Tocher approximation.

## 1. Introduction

The numerous applications as well as the significant statistical properties of the normal distribution make it one of the most important continuous distribution functions. Usually, researchers in various areas including but not limited to medical, engineering, social studies use the cumulative distribution function of normal distribution in testing and verifying various problems and conjectures in these fields.

It is well known that a random variable $Z$ is normally distributed with mean 0 and standard deviation equals 1 if the probability density function of $Z$ is given by,

$$f(z) = \frac{1}{\sqrt{2\pi}} e^{\frac{-z^2}{2}}, \quad -\infty < z < \infty.$$

Accordingly, the cumulative distribution function (CDF) of $Z$ is,

$$\Phi(z) = \frac{1}{\sqrt{2\pi}} \int_{-\infty}^{z} e^{\frac{-t^2}{2}} dt.$$

Clearly, there is no closed form solution for the CDF of the normal distribution and this is one of the most important challenges to be discussed by researchers. For this reason, many approximations of $\Phi(z)$ have been proposed in the literature oover the last six decades. Recently, Eidous and Abu Shareefa (2020) and the references therein revised and discussed nearly 45 existing approximations of the normal CDF and suggested another 9 additional approximations of $\Phi(z)$.

Given that $\widehat{\Phi}(z)$ is an approximation of $\Phi(z)$ evaluated at a specific value of $z$, we plan to use two well-known criteria to measure the accuracy of $\widehat{\Phi}(z)$. The first measure is the maximum absolute error ($MXAE$) of $\widehat{\Phi}(z)$, which is defined as follows,

$$MXAE\left(\widehat{\Phi}(z)\right) = \max_{z} \left|\widehat{\Phi}(z) - \Phi(z)\right|.$$

The second measure of accuracy is the mean absolute error ($MAE$) of $\widehat{\Phi}(z)$, which is given by,

$$MAE\left(\widehat{\Phi}(z)\right) = \frac{\sum_{z} \left|\widehat{\Phi}(z) - \Phi(z)\right|}{\# \, of \, z},$$

where $n$ is the number of $z$ values selected for a given domain of interest. It is noteworthy that we compute both $MXAE$ and $MAE$ for any number of selected values of $z$. Therefore, we plan to choose the support values of $z$ between 0 to 4 with step 0.01, which means that the above accuracy measures will be evaluated at 401 points.

In this article, we discuss some existing approximations of $\Phi(z)$ that have been introduced in the literature based on Tocher's approximation of $\Phi(z)$ (see Lin, 1990; Divgi, 1990; Vedder, 1993 and Boiroju and Rao, 2014). We aim at proposing new approximations of $\Phi(z)$ to improve the existing Tocher's approximation. We intend to use the previously mentioned accuracy measures $MXAE$ and $MAE$ to investigate and compare the performance of the new proposed approximation as well as some of the existing approximations.

## 2. Existing approximations of $\Phi(z)$

A pioneer and simple idea was introduced by Tocher (1963) to present a closed form approximation of $\Phi(z)$ as follows,

$$\Phi_1(z) = \frac{e^{2\sqrt{\frac{2}{\pi}z}}}{1+e^{2\sqrt{\frac{2}{\pi}z}}},$$

$$= \frac{1}{1+e^{-2\sqrt{\frac{2}{\pi}z}}}, \qquad z > 0$$

Note that if $z < 0$, we may use the symmetry property of the normal distribution via the relation $\Phi(z) = 1 - \Phi(-z)$. The $MXAE$ based on $\Phi_1(z)$ appears to be $1.77 \times 10^{-2}$.

Several approximations have been suggested in the literature to improve the accuracy of Tocher's approximation and we list some of these approximations in the sequel:

1. The first improvement of Tocher's approximation considered in this article is proposed by Lin (1990) as follows:

$$\Phi_2(z) = \frac{1}{1+e^{-\frac{4.2\pi z}{9-z}}}, \qquad 0 \leq z < 9$$

Which produces $MXAE$ to be equal to $6.69 \times 10^{-3}$.

2. The second improvement is proposed by Divgi (1990) as follows:

$$\Phi_3(z) = \frac{1}{1+e^{-y}},$$

where $y = 1.526\, z\, (1 + 0.1034\, z)$. The $MXAE$ of $\Phi_3(z)$ is $2.10 \times 10^{-3}$.

3. The third approximation is proposed by Vedder (1993) as follows

$$\Phi_4(z) = \frac{1}{1+e^{-y}},$$

where $y = \sqrt{8/\pi}z + \sqrt{2/\pi}(4-\pi)z^3/3\pi$. In this case, the $MXAE$ of $\Phi_4(z)$ equals to $3.14 \times 10^{-4}$.

4. The fourth improvement is introduced by Waissi and Rossin (1996) in the following form

$$\Phi_5(z) = \frac{1}{1+e^{-y}},$$

where $y = \sqrt{\pi}\,(0.9z + 0.0418198z^3 - 0.0004406z^5)$. The $MXAE$ of $\Phi_5(z)$ is $4.37 \times 10^{-5}$.

5. Bowling *et al.* (2009) proposed another improvement following the same previously mentioned ideas as

$$\Phi_6(z) = \frac{1}{1+e^{-1.5976\,z - 0.07056\,z^3}},$$

with $MXAE$ equals to $1.42 \times 10^{-4}$.

6. The next approximation considered in this article was proposed by Boiroju and Rao (2014) given as

$$\Phi_7(z) = \frac{1}{1+e^{-y}},$$

where

$$y = \frac{1}{2}(-0.506445 + 10.4467 \tanh(1.3448 + 0.3264z)$$
$$+ 9.8475 \tanh(-1.3519 + 0.3376z) + 1.5976\,z$$
$$+ 0.070565992\,z^3).$$

In this case, the value of $MXAE$ corresponding to $\Phi_7(z)$ is $2.41 \times 10^{-5}$.

7. Finally, Eidous and Ananbeh (2021) proposed another improvement to enhance the accuracy level of the approximation as follows

$$\Phi_8(z) = \frac{1}{1+e^{-y}},$$

where

$$y = 1.5957764z + 0.0726161\,z^3 + 0.00003318\,z^6 - 0.00021785\,z^7$$
$$+ .00006293\,z^8 - 0.00000519\,z^9$$

The $MXAE$ of $\Phi_8(z)$ obtained for this approximations is $7.62 \times 10^{-7}$.

## 3. Proposed Approximation of $\Phi(x)$

In this section, we propose a new approximation to improve the accuracy of Tocher's approximation and the accuracy of previous approximations as well. The proposed approximation of $\Phi(x)$ is given as

$$\Phi_9(z) = \frac{1}{1+e^{-az}}, \quad z \geq 0$$

where

$a = 1.5957691187 + 5.37366 \times 10^{-8}\,z + 0.72670769\,z^2 - 9.229 \times 10^{-7}\,z^3 - 5.3498 \times 10^{-5}\,z^4 - 9.0342 \times 10^{-5}\,z^5 + 1.049448 \times 10^{-4}\,z^6 - 3.0263611 \times 10^{-3}\,z^7 + 2.99472642 \times 10^{-4}\,z^8 - 1.98173433 \times 10^{-4}\,z^9 + 9.4285766 \times 10^{-5}\,z^{10} - 3.1366467 \times 10^{-5}\,z^{11} + 7.1524366 \times 10^{-6}\,z^{12} + 1.09550613 \times 10^{-6}\,z^{13} + 1.079959 \times 10^{-7}\,z^{14} - 6.208087 \times 10^{-9}\,z^{15} + 1.585371 \times 10^{-10}\,z^{16}.$

This approximation may be written as

$$a = \sum_{j=1}^{17} k_j z^{j-1},$$

where $k_j, j = 1,2,3,\ldots,17$ are given in Table 1 below.

Figure 1 below illustrates the differences between our proposed approximation and the true value of $\Phi(z)$. A thorough look allows us to see that the $MXAE$ value of the approximation $\Phi_9(z)$ is $4.43429 \times 10^{-10}$ which occurs at $z = 0.794634$.

**Table 1:** The values of $k_j, j = 1,2,3,\ldots,17$.

|  |  |
|---|---|
| $k_1$ | 1.5957691187 |
| $k_2$ | $5.37366 \times 10^{-8}$ |
| $k_3$ | 0.72670769 |
| $k_4$ | $-9.229 \times 10^{-7}$ |
| $k_5$ | $5.3498 \times 10^{-5}$ |
| $k_6$ | $-9.0342 \times 10^{-5}$ |
| $k_7$ | $1.049448 \times 10^{-4}$ |
| $k_8$ | $-3.0263611 \times 10^{-3}$ |
| $k_9$ | $2.99472642 \times 10^{-4}$ |
| $k_{10}$ | $-1.98173433 \times 10^{-4}$ |
| $k_{11}$ | $9.4285766 \times 10^{-5}$ |
| $k_{12}$ | $-3.1366467 \times 10^{-5}$ |
| $k_{13}$ | $7.1524366 \times 10^{-6}$ |
| $k_{14}$ | $1.09550613 \times 10^{-6}$ |
| $k_{15}$ | $1.079959 \times 10^{-7}$ |
| $k_{16}$ | $-6.208087 \times 10^{-9}$ |
| $k_{17}$ | $1.585371 \times 10^{-10}$ |

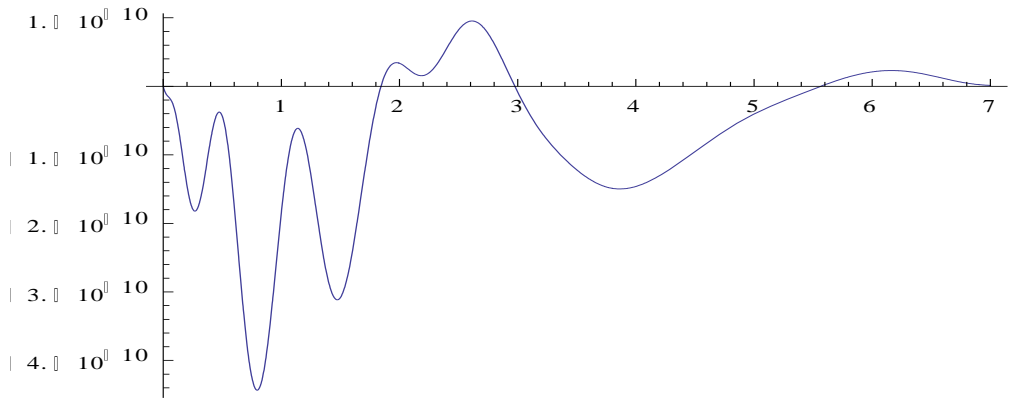

**Figure 1:** The difference between $\Phi_9(z)$ and $\Phi(z)$.

## 4. Approximation of the Inverse of $\Phi(z)$

Let $p = \Phi(z)$ be the cumulative distribution function of standard normal, where $z \geq 0$. The inverse of the cumulative distribution function of the standard normal distribution is $z = \Phi^{-1}(p)$. Assuming that the value of $p$ is given, we are interested in obtaining the approximate value of $z$ ($z \geq 0$) at which the area underneath it is $p$ under the standard normal curve.

Schmeiser (1979) derived a simple approximation for $z$, which is given by,

$$\hat{z}_1 = \frac{(p^{0.135} - (1-p)^{0.135})}{0.1975}, \qquad p \geq 0.5.$$

Later, Shore (1982) derived the following approximation of $z$ for a given value of $p \geq 0.5$,

$$\hat{z}_2 = -5.531\left(\left(\frac{1-p}{p}\right)^{0.1193} - 1\right).$$

In this section, we introduce a new approximation of $z$. Based on the Polya approximation of $p = \Phi(z)$ (Polya,1949), which is given by,

$$p = \Phi_P(z) = 0.5\left(1 + \sqrt{1 - e^{-\left(\frac{2}{\pi}\right)z^2}}\right).$$

We suggest the following approximation of $z$,

$$\hat{z}_3 = \sqrt{-\frac{1}{d_1}\log(1 - [2(p-0.5)]^2)},$$

where
$$d_1 = 0.8039 - 0.9446\,p + 1.5806\,p^2 - 1.7824\,p^4 + 1.5098\,p^6 - 0.5689\,p^8.$$
If we define $\Delta_i = \hat{Z}_i - Z$, then we can easily see the changes in $\Delta_i$ with respect to the values of $p$. Figure 2 illustrates the behavior of $\Delta_3$ that reflects that accuracy of our proposed approximation.

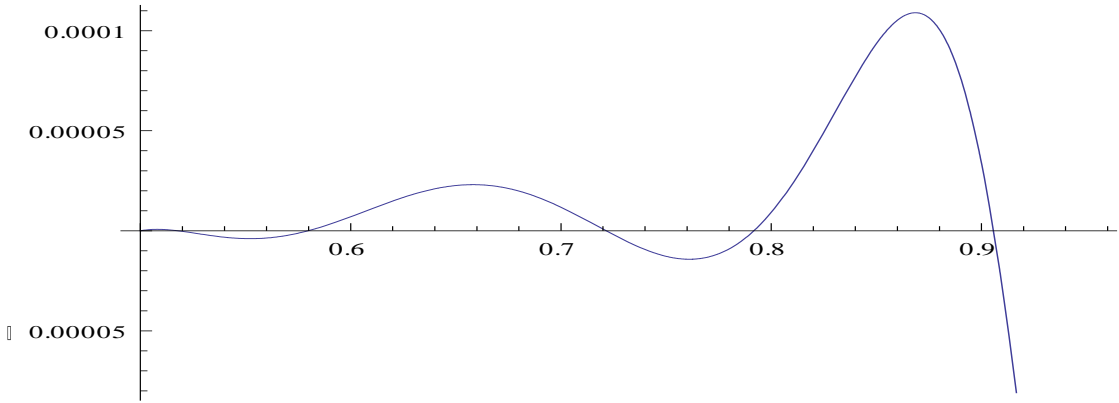

**Figure 2:** Graphical presentation of $\Delta_3 = \hat{z}_3 - z$ with respect to $p$.

## 5. Comparisons and Conclusions

To shed more light on the accuracy of variant approximations of $\Phi(z)$ that discussed in this article, we obtain the values of $MAE$ for each approximation depending on grid values of $z$ starting from 0 to 5 with step 0.001 and we summarize these results in Table 2. To simplify the comparison mechanism, we include, in Table 2, the $MXAE$ values of the approximations discussed in this article.

Table 2 clearly shows that the proposed approximation $\Phi_9(z)$ is more accurate than the other approximations based on the two criteria $MXAE$ and $MAE$.

The other part of our computational aspect focuses on investigating the accuracy of the proposed approximation $\hat{z}_3$. A number of values for z are selected from 0 to 4.8 with step 0.4. For each value of z, we compute the corresponding value of $p = \Phi(z)$ and then we obtain the proposed estimate ($\hat{z}_3$). For sake of comparison, we include the results of the two approximations $\hat{z}_1$ and $\hat{z}_2$ in Table 3 and Table 4.

This allows the reader to compare the values of the different approximations with the exact values of z (first column). To simplify the comparison process, we provided the differences $\Delta_i = \hat{z}_i - z$ for $i = 1, 2, 3$ in Table 4.

Table 3 clearly shows that the proposed approximation $\hat{z}_3$ is more accurate than $\hat{z}_1$ and $\hat{z}_2$ especially when the exact value of $\Phi(z) < 0.99$. Table 4 emphasizes on the high quality of the proposed estimator in view of the values of $\Delta_3$.

All results and graphs are obtained by using Mathematica, Version 12.

**Table 2**: The $MXAE$ and the $MAE$ for the approximations $\Phi_1(z)$ to $\Phi_9(z)$.

| Approximation | $MXAE$ | $MAE$ |
| --- | --- | --- |
| $\Phi_1(z)$ | $1.77 \times 10^{-2}$ | $\mathbf{7.05 \times 10^{-3}}$ |
| $\Phi_2(z)$ | $6.69 \times 10^{-3}$ | $\mathbf{1.10 \times 10^{-3}}$ |
| $\Phi_3(z)$ | $2.10 \times 10^{-3}$ | $\mathbf{9.78 \times 10^{-4}}$ |
| $\Phi_4(z)$ | $3.14 \times 10^{-4}$ | $\mathbf{9.99 \times 10^{-5}}$ |
| $\Phi_5(z)$ | $4.37 \times 10^{-5}$ | $\mathbf{1.69 \times 10^{-5}}$ |
| $\Phi_6(z)$ | $1.42 \times 10^{-4}$ | $\mathbf{6.88 \times 10^{-5}}$ |
| $\Phi_7(z)$ | $2.41 \times 10^{-5}$ | $\mathbf{7.26 \times 10^{-6}}$ |
| $\Phi_8(z)$ | $7.62 \times 10^{-7}$ | $\mathbf{1.82 \times 10^{-7}}$ |
| $\Phi_9(z)$ | $4.43 \times 10^{-10}$ | $\mathbf{9.62 \times 10^{-11}}$ |

## 6. Discussion

In this article, we focused on developing a new approximation of the cumulative normal distribution $\Phi(z)$ for given values of $z$. In addition, we suggested an invertible approximation of $z$ for given values of $p = \Phi(z)$. Accordingly, it is found that the performance of the proposed approximation is dominant when compared to some of the existing approximations for a wide range of true values of $z$.

**Table 3:** The three approximations of z obtained at selected true values of z corresponding to $p$.

| z | $p = \Phi(z)$ | $\hat{z}_1$ | $\hat{z}_2$ | $\hat{z}_3$ |
|---|---|---|---|---|
| 0.0 | 0.5000 | 0.000 | 0.000 | 0.000 |
| 0.4 | 0.6554 | 0.3976 | 0.4084 | 0.4000 |
| 0.8 | 0.7881 | 0.7969 | 0.8024 | 0.8000 |
| 1.2 | 0.8849 | 1.1989 | 1.1948 | 1.1999 |
| 1.6 | 0.9452 | 1.6038 | 1.5932 | 1.6003 |
| 2.0 | 0.9773 | 2.0093 | 1.9993 | 1.9975 |
| 2.4 | 0.9918 | 2.4105 | 2.4097 | 2.3864 |
| 2.8 | 0.9974 | 2.7999 | 2.8168 | 2.7660 |
| 3.2 | 0.9993 | 3.1686 | 3.2109 | 3.1386 |
| 3.6 | 0.9998 | 3.5084 | 3.5826 | 3.5068 |
| 4.0 | 0.99997 | 3.8130 | 3.9239 | 3.8725 |
| 4.4 | 0.99999 | 4.0783 | 4.2293 | 4.2366 |
| 4.8 | 1.0000 | 4.3032 | 4.4958 | 4.5997 |

**Table 4:** The differences $\Delta_i = \hat{z}_i - z, i = 1, 2, 3$ corresponding to the approximation of z evaluated at selected exact values of z.

| z | $p = \Phi(z)$ | $\Delta_1$ | $\Delta_2$ | $\Delta_3$ |
|---|---|---|---|---|
| 0.0 | 0.5000 | 0. | 0. | 0. |
| 0.4 | 0.6554 | -0.00238 | 0.00839 | -0.00002 |
| 0.8 | 0.7881 | -0.00313 | 0.00237 | $3.28071 \times 10^{-6}$ |
| 1.2 | 0.8849 | -0.00107 | -0.00521 | -0.00009 |
| 1.6 | 0.9452 | 0.00380 | -0.00685 | 0.00025 |
| 2.0 | 0.9773 | 0.00932 | -0.00070 | -0.00025 |
| 2.4 | 0.9918 | 0.01051 | 0.00972 | -0.01360 |
| 2.8 | 0.9974 | -0.00015 | 0.01681 | -0.03403 |
| 3.2 | 0.9993 | -0.03141 | 0.01094 | -0.06142 |
| 3.6 | 0.9998 | -0.09157 | -0.01738 | -0.09318 |
| 4.0 | 0.99997 | -0.18706 | -0.07607 | -0.12752 |
| 4.4 | 0.99999 | -0.32173 | -0.17066 | -0.16342 |
| 4.8 | 1.0000 | -0.49679 | -0.30416 | -0.20026 |